\documentclass[10pt, journal,fleqn]{ieeeconf}  

\IEEEoverridecommandlockouts     

\usepackage{color,amsmath,amstext,amsfonts,amssymb, mathrsfs,mathtools}
\usepackage{graphicx,algorithm,algorithmic}
\usepackage{tikz}
\usepackage{lipsum,adjustbox}
\usepackage{setspace}
\usepackage{array} 
\usepackage{booktabs}  
\usepackage{bbm}
\usepackage{wasysym}
\usepackage{textcomp}
\usepackage{hyperref} 
\usepackage[sort,compress]{cite}
\usepackage{subcaption} 

\renewcommand{\thesubsubsection}{{\it \Alph{subsection}.\arabic{subsubsection}.}}

\newcounter{l1}
\newcounter{l2}
\newcounter{l3}
\newcommand{\bdotlist}{\begin{list}{$\bullet$}{}}
\newcommand{\bboxlist}{\begin{list}{$\Box$}{}}
\newcommand{\bbboxlist}{\begin{list}{\raisebox{.005in}{{\tiny $\blacksquare$ \ \ }}}{}}
\newcommand{\bdashlist}{\begin{list}{$-$}{} }
\newcommand{\blist}{\begin{list}{}{} }
\newcommand{\barablist}{\begin{list}{\arabic{l1}}{\usecounter{l1}}}
\newcommand{\balphlist}{\begin{list}{(\alph{l2})}{\usecounter{l2}}}
\newcommand{\bAlphlist}{\begin{list}{\Alph{l2}.}{\usecounter{l2}}}
\newcommand{\bdiamlist}{\begin{list}{$\diamond$}{}}
\newcommand{\bromalist}{\begin{list}{(\roman{l3})}{\usecounter{l3}}}

\providecommand{\norm}[1]{\lVert#1\rVert}

\newcommand{\beq}{\begin{equation}}
\newcommand{\eeq}{\end{equation}}

\title{\LARGE \bf
Improved Attention Models for Memory Augmented \\ Neural Network Adaptive Controllers
}

\author{Deepan Muthirayan, Scott Nivison and Pramod P. Khargonekar  
\thanks{Deepan Muthirayan and Pramod P. Khargonekar are with the Department of Electrical Engineering and Computer Sciences, University of California, Irvine, CA 92697. Email: \{dmuthira, pramod.khargonekar\}@uci.edu}
\thanks{Scott Nivison is with the Munitions Directorate, Air Force Research Laboratory, Eglin AFB, FL, USA. Email:  {scott.nivison}@us.af.mil.}
\thanks{Supported in part by the National Science Foundation under Grant Number ECCS-1839429.}
}

\begin{document}
\maketitle

\thispagestyle{empty}
\pagestyle{empty} 

\begin{abstract}
We introduced a {\it working memory} augmented adaptive controller in our recent work. The controller uses attention to read from and write to the working memory. Attention allows the controller to read specific information that is relevant and update its working memory with information based on its relevance, similar to how humans pick relevant information from the enormous amount of information that is received through various senses. The retrieved information is used to modify the final control input computed by the controller. We showed that this modification speeds up learning.

In the above work, we used a soft-attention mechanism for the adaptive controller. Controllers that use soft attention update and read information from all memory locations at all the times, the extent of which is determined by their relevance. But, for the same reason, the information stored in the memory can be lost. In contrast, hard attention updates and reads from only one location at any point of time, which allows the memory to retain information stored in other locations. The downside is that the controller can fail to shift attention when the information in the current location becomes less relevant.

We propose an attention mechanism that comprises of (i) a hard attention mechanism and additionally (ii) an attention reallocation mechanism. 
The attention reallocation enables the controller to reallocate attention to a different location when the relevance of the location it is reading from diminishes. The reallocation also ensures that the information stored in the memory before the shift in attention is retained which can be lost in both soft and hard attention mechanisms. 
Through detailed simulations of various scenarios for two link robot robot arm systems we illustrate the effectiveness of the proposed attention mechanism.
\end{abstract}

\section{Introduction}

Even though there is remarkable progress in machine learning, robotics and autonomous systems, humans still outperform intelligent machines in a wide variety of tasks and situations \cite{lake2017building}. Central to the functioning of human cognitive system are several functions like perception, memory, attention, reasoning, problem solving, thinking and creativity. Thus, it is only natural to ask, can functions inspired from human cognition improve control algorithms? With this neuro-cognitive science inspiration, we recently introduced the concept of memory augmented neural adaptive controllers in \cite{muthirayan2019memory, muthirayan2019memorynonlinear}. In this paper, we focus on the notion of attention in this setting. 

{\it Attention} is the state of focused awareness on some aspects of the environment. Attention allows humans to focus on sensory information that are relevant and essential at a particular point of time. This is especially important considering, at any moment of time, the human takes in enormous amount of information from visual, auditory, olfactory, tactile and taste senses. The human-inspired attention mechanism have been instrumental in machine learning applications such as machine translation \cite{vaswani2017attention}. Attention based models have also improved deep learning models like neural networks \cite{wang2018non}, reinforcement learning (RL) \cite{graves2016hybrid} and are the state-of-the-art in meta learning \cite{mishra2017simple} and generative adversarial networks \cite{zhang2018self}. We believe that the idea of attention holds tremendous potential for learning in cyber-physical systems.

Inspired by human attention and the recent successes in deep learning, we explore attention models for control algorithms in a specific context. We consider the working memory augmented NN adaptive controllers proposed in our recent work \cite{muthirayan2019memory}, \cite{muthirayan2019memorynonlinear}. Here, the controller stores and retrieves specific information to modify its final control input. We showed that this modification speeds up the response of the controller to abrupt changes. Attention is relevant here because, the main controller uses an attention mechanism to read and write to the working memory. The natural question to ask is what is a good attention mechanism for such an application? 

Our previous work \cite{muthirayan2019memory} used a specific form of attention called soft-attention, where the central controller reads and writes to all locations in the working memory. Here, the extent to which the memory contents are erased and rewritten or contribute to the final read output depends on the relevance of the content at a particular location in the memory. While soft attention mechanisms can be effective by retrieving relevant information from all locations, the same feature can lead to loss of information.

In contrast, hard attention mechanisms \cite{nivison2018sparse} are mechanisms that read and write to only one location at any point of time. A controller that uses hard attention does not lose the information stored in other locations since at any point of time only one location is read from or updated. The disadvantage is that the controller can fail to shift attention when the current information becomes less relevant. In addition, this information can be lost due to continual modifications.

In this paper, we propose an attention mechanism that is a combination of {\it hard attention} and an {\it attention reallocation} mechanism. The attention reallocation mechanism allows the controller to shift attention to a different location when the current location becomes less relevant. This also allows the memory to retain the information stored in it before the shift was forced by the reallocation mechanism. Thus, the mechanism we propose can overcome the limitations of prior hard and soft attention mechanisms. 

The setting we consider is a well-studied NN adaptive control setting. This setting comprises of an unknown nonlinear function that is to be compensated. Typically, the neural network is used to directly compensate the unknown function. The literature on NN based adaptive control is extensive \cite{narendra1990identification, yecsildirek1995feedback, lewis1996multilayer, narendra1997adaptive, kwan2000robust, calise2001adaptive, chen2001nonlinear, ge2004adaptive}. In the setting here, we consider nonlinear uncertainties that can vary with time including variations that are {\it abrupt} or {\it sudden}. The objective for the controller is to adapt quickly even after such abrupt changes. We make the assumptions that (i) the abrupt changes are not large and (ii) the system state is observable. 

In Section \ref{sec:cont-arch}, we revisit the {\it Memory Augmented Neural Network} (MANN) adaptive controller proposed in our recent works \cite{muthirayan2019memory}, \cite{muthirayan2019memorynonlinear}. In Section \ref{sec:seg-mem-int}, we discuss the working memory interface and the proposed attention mechanism. Finally, in Section \ref{sec:disc}, we provide a detailed discussion substantiating the improvements in learning obtained by using the attention mechanisms proposed in this paper. We do not include the stability proof for the controller described here. The proof outlined in \cite{muthirayan2019memory} can be trivially extended to the closed loop system discussed here.  

\section{Control Architecture}
\label{sec:cont-arch}

In this section, we briefly introduce the memory augmented control architecture for adaptive control of continuous time systems proposed in our earlier work \cite{muthirayan2019memory}. The central inspiration behind this idea is the working memory in the human memory system.

\begin{figure}
\center
\begin{tikzpicture}[scale = 0.45]

\draw [draw=blue, fill=blue, fill opacity = 0.1, rounded corners, thick] (-2, 4) rectangle (2, 5);
\draw (0, 4.5) node[align=center] {\tiny $f$};

\draw [draw=blue, fill=blue, fill opacity = 0.1, rounded corners, thick] (-2, 1) rectangle (2, 3);
\draw (0, 2) node[align=center] {\tiny Plant};
\draw  [->, thick] (-2,-0.5) -- (-3,-0.5) -- (-3,1.5) -- (-2,1.5);
\draw  [->, thick] (2,1.5) -- (3,1.5) -- (3,-0.5) -- (2,-0.5);
\draw  [->, thick] (-2,4.5) -- (-3,4.5) -- (-3,2.5) -- (-2,2.5);
\draw  [->, thick] (2,2.5) -- (3,2.5) -- (3,4.5) -- (2,4.5) ;
\draw (-3.25, 1.5) node[align=center] {\small $u$};
\draw (-3.25, 2.5) node[align=center] {\small $w$};
\draw (3.25, 1.5) node[align=center] {\small $x$};
\draw (3.25, 2.5) node[align=center] {\small $z$};

\draw [draw=blue, fill=blue, fill opacity = 0.1, rounded corners, thick] (-2, 0) rectangle (2, -1);
\draw (0, -0.5) node[align=center] {\tiny Controller};
\draw  [<->, thick] (0,-1) -- (0,-2);

\draw [draw=blue, fill=blue, fill opacity = 0.1, rounded corners, thick] (-2, -2) rectangle (2, -3);
\draw (0, -2.5) node[align=center] {\tiny Working Memory};

\draw [draw=blue, fill=blue, fill opacity = 0.1, rounded corners, thick] (5, 0.25) rectangle (9.25, 4);
\draw [draw=blue, fill=blue, fill opacity = 0.1, rounded corners, thick] (7.75, 1) rectangle (9, 3);
\draw (8.4,2) node[align=center] {\tiny NN};
\draw (6,2) node[circle , draw, thick] (b) {\tiny $+$};
\draw [->, thick] (b) -- (4.5,2);
\draw [->, thick] (7.75,2) -- (b);
\draw [->, thick] (6,3.5) -- (b);
\draw [->, thick] (6,0.5) -- (b);
\draw (7, 4.25) node[align=center] {\tiny Neural Adaptive Controller};
\draw (4.25,2) node[align=center] {\small $u$};
\draw (6,3.75) node[align=center] {\small $v$};
\draw (6.5,0.75) node[align=center] {\small $u_{bl}$};
\draw (7.3,2.25) node[align=center] {\small $u_{ad}$};

\draw [draw=blue, fill=blue, fill opacity = 0.1, rounded corners, thick] (5, -2) rectangle (9.25, -3);
\draw (7, -2.5) node[align=center] {\tiny Working Memory};
\draw  [->, thick] (6.5,-2) -- (6,0.25);
\draw  [->, thick]  (8.25,0.25) -- (7.75,-2);
\draw (5.75,-1) node[align=left] {\tiny Read};
\draw (8.5,-1) node[align=right] {\tiny Write};

\end{tikzpicture}
\caption{Left: general controller augmented with working memory. Right: memory augmented neural adaptive controller. $f$ is the uncertainty.}
\label{fig:cwmem}
\end{figure}
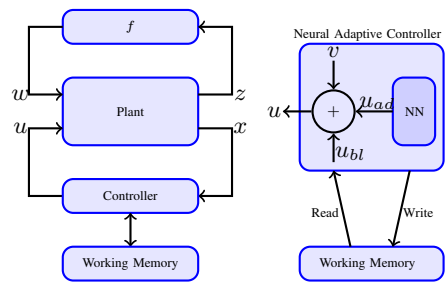

{\it Proposed Architecture}: The general architecture is depicted in Fig. \ref{fig:cwmem}. The proposed architecture augments an external working memory to the general dynamic feedback controller. In our previous work, we specialized it to the specific controller that augments an external working memory to a neural network. In this paper, we focus on the attention models for the working memory. We propose several attention mechanisms for the controller. The novelty of our proposal lies in how the controller can reallocate its attention under certain circumstances. 

{\it Specialization to NN Adaptive Control}: in neural adaptive control, control input $u$ to the plant is computed based on state feedback, error feedback and the NN output. The control input is a combination of base controller $u_{bl}$, which is problem specific, the NN output $u_{ad}$ and a ``robustifying term'' $v$ \cite{lewis1996multilayer, lewis1998neural}. The final control input is given by
\beq u = u_{bl} + u_{ad} + v. \label{eq:claw} \eeq

Since, it is assumed that the system state is observable, the output of the plant is the system state. The control term $u_{bl}$ is computed based on the output of an error evaluator, for example, the error between the system output and the desired trajectory in a trajectory tracking problem. The control term $u_{ad}$ is the final NN output. This output is typically used to compensate an unknown nonlinear function in the system dynamics. The robustifying term is introduced to compensate the higher order terms that are left out in the compensation of the unknown function by the term $u_{ad}$.

In the memory augmented NN adaptive control we proposed in \cite{muthirayan2019memory}, the working memory acts as a complementing memory system to the NN. The output of the NN block, $u_{ad}$ is computed by combining the information read from the external working memory and the NN. The exact form of this term and how it is modified based on the contents of the working memory is described in Section \ref{sec:seg-mem-int}. We showed in \cite{muthirayan2019memory} that this modification speeds up the response of the closed loop system to abrupt changes.

{\it Notation}: The system state is denoted by $x \in \mathbb{R}^n$. The estimated NN weight matrices are given by $\hat{W}, \hat{V}, \hat{b}_w$ and $\hat{b}_v$. 
We denote the NN used to approximate the function $f(x)$ by $\hat{f}$, which is given by $\hat{f}= \hat{W}^T\sigma(\hat{V}^Tx + \hat{b}_v) + \hat{b}_w$. 

We introduce a function called $\text{softmax(.)}$, takes in as input a vector $a$ and outputs a vector of the same length. The $i$th component of $\text{softmax(.)}$ is given by
\beq \text{softmax(a)}_i  = \frac{\exp{(a_i)}}{\sum_j \exp{(a_j)}}. \eeq
We introduce two other vector functions which appear in the NN update laws. We denote these functions by $\hat{\sigma}$ and $\hat{\sigma}'$ which are defined by:
\[ \hat{\sigma}  = \left[ \begin{array} {c} \sigma(\hat{V}^Tx + \hat{b}_v) \\ 1 \end{array} \right] \nonumber \]
\beq \hat{\sigma}' = \left[ \begin{array} {c} \text{diag}(\sigma(\hat{V}^Tx + \hat{b}_v)\odot(1 - \sigma(\hat{V}^Tx + \hat{b}_v))) \\ \mathbf{0}^T \end{array} \right] \eeq
where $\mathbf{0}$ is a zero vector of dimension equal to the number of hidden layer neurons. 

\section{Working Memory and Attention}
\label{sec:seg-mem-int}

In this section, we first summarize the two working memory operations, i.e., Memory Write and Memory Read, introduced in our previous works \cite{muthirayan2019memory}, \cite{muthirayan2019memorynonlinear}. We then provide a detailed discussion on the various attention mechanisms and the attention reallocation mechanism. Finally, we discuss how the Memory Read output is used to modify the final output $u_{ad}$.

\subsection{\it Memory Write:} The Memory Write equation for the working memory is
\begin{equation}
 \dot{h}_{i} = -w_r(i)h_{i} + c_w w_r(i) h_w + w_r(i)\hat{W} h^T_{e},
 \label{eq:memorywrite} 
\end{equation}
where $h_w$ is the write vector, $w_r(i)$ is the factor that determines whether the memory vector $i$ is eligible for update or not and $c_w$ is a constant. The write vector $h_w$ corresponds to the new information that can be used to update the contents of the memory. The write vector $h_w$ for this interface is given by,
\beq h_w = \sigma(V^T \tilde{x} + \hat{b}_{v}). \label{eq:writevec} \eeq

Specifying the write vector to be the current hidden layer value allows the memory to store and reuse them later to modify the final NN output. 

The factor $w_r(i)$s determine the elegibility of the respective locations for update and are determined by an addressing mechanism. The controller generates a query $q$ which could be, for example, the current state or the current hidden layer output itself. Each memory location $i$ is typically associated with a key $k_i$, which serve as its identifier. The keys $k_i$s are compared with the generated query to deterimine the factors $w_r(i)$s. The factor $w_r(i)$ is set to be $1$ for the $i$ whose $k_i$ is closest to the query $q$ and the rest are set to zero. The location whose corresponding factor is one becomes eligible and is updated. In machine learning parlance this type of an addressing mechanism is referred to as {\it hard attention} \cite{luong2015effective}. 

Later, we discuss several attention mechanisms that differ based on how the keys are determined and how the query is specified. 

\subsection{\it Memory Read:} The Memory Read output, $h_o$, for the interface is given by
\beq \ h_{o} = h w_r, \label{eq:memoryread} \eeq

where $w_r(i)$s are the same set of factors discussed earlier. Essentially, the Memory Read output is a weighted combination of the respective memory contents with the weights being $w_r(i)$s. The factors $w_r(i)$s are a natural choice for the weights because they have a direct correspondence to the relevance of the information in their respective locations. 

\subsection{Attention Mechanism}
\label{sec:addmech}

%
We propose an attention mechanism that comprises of (i) a hard attention and (ii) an attention reallocation mechanism. The attention reallocation mechanism shifts the attention to a different location when the relevance of the information in the current location diminishes. After the shift, the information in the previous location is retained because in the period before the next shift the proposed mechanism uses hard attention. Mechanisms that use hard attention without attention reallocation can fail to shift attention and continue to read from and modify this information, eventually losing this information. Thus, attention reallocation plays a complementary role to hard attention. Below, we discuss two different hard attention mechanisms for the proposed controller (i) where the key is state based and (ii) where the key is representation based. We then discuss the attention reallocation mechanism.

\subsubsection{State based Hard Attention}
\label{sec:dyn-state-key}

In this design, the keys are specified to be a set of points in the state space. The update equations for the keys are specifed to be an asymptotically stable first order dynamic system whose state is the key vector $k_i$ and input is the sub-vector of the state $\underbar{x}$. Thus, the key update equations are given by:
\beq \dot{k}_i  = - c_k w_r(i) (k_i - \underbar{x}), \label{eq:keyupdynst} \eeq
where the constant $c_k$ is a design constant. The constant $c_k$ determines the response time of this equation and is set such that the final output of this equation is a good representation of the states visited when its corresponding memory location was updated. 

The query needs to be set such that the controller can retrieve values relevant to the current state of the system. Given that the keys are points in the state space, we set the query $q$ to be the current state itself, i.e., 
\beq q = \underbar{x}.\eeq 
In the simulation examples that we discuss later $\underbar{x}$ is chosen to be the position variables in the state vector. 

A hard attention mechanism selects the location whose key is closest to the query. In this design, this location is determined by
\beq i^* = \text{argmin}_i \norm{q-k_i}_{\infty}. \label{eq:indsel} \eeq
Given $i^{*}$, the factors $w_r(i)$s are naturally given by
\beq w_r(i) = \left\{ \begin{array}{cc} 1, &  \text{if} \ i = i^*, \\ 0, & \text{otherwise}. \end{array} \right. \label{eq:factdef} \eeq

\subsubsection{Representation based Hard Attention}
\label{sec:dyn-rep-key}

Alternatively, we can specify the keys as a set of points in the hidden layer feature space of the neural network (NN). This is a reasonable approach because the hidden layer features by definition are a representation of the NN input space. 

In this design, the key for the respective memory locations are given by
\beq k_i = h_i. \label{eq:keyupdynrep} \eeq 

We choose the memory vectors themselves as the keys for the following reasons. Firstly, the memory vectors provide a dynamic set of points in the feature space. Secondly, a key by design should contain information about the memory content and the scenario it represents and the memory vectors satisfy this criterion. The keys for the respective locations are formally given by

The query $q$ should be such that the controller can retrieve information from the memory that is relevant. Assuming that the current scenario and the scenario that the memory contents correspond to are not very different, the information in the location that is likely to be relevant is the location whose key is closest to the current hidden layer output. The controller can retrieve this information by specifying the query $q$ to be the current hidden layer output of the NN itself. Hence, we define $q$ as:
\beq q = \sigma(V^T\tilde{x}+b_v). \eeq
 
In this design, the location that is selected by the hard attention mechanism is given by
\beq i^* = \text{argmin}_i \norm{q-1/c_w k_i}_{\infty}. \label{eq:indsel-1} \eeq 

The factor $1/c_w$ in the above equation accounts for the same factor in the Memory Write equation \eqref{eq:memorywrite} which is also the key update equation in this case. Having defined $i^*$, the factor $w_r(i)$s are given by the same equation \eqref{eq:factdef}.

\subsection{Attention Reallocation}

The attention reallocation mechanism that we propose continually checks whether the content of any of the memory locations is nearer to the current hidden layer output. If, at some point, the hidden layer output deviates from any of the contents beyond this threshold $\theta$ then the controller reallocates attention to the least relevant location with its value re-initialized at the current hidden layer value. Such a design is likely to ensure that, at any point of time, there is at least one memory location whose content is `very' relevant to the current scenario. 

Attention reallocation also ensures that the memory does not forget the information stored before the shift in attention. This is because once the shifts occurs, the location where the information is stored is neither updated nor read, at least for a certain period, until the attention shifts back to this location. 
We note that, in both hard and soft attention, this information is likely to be lost.

The decision to shift attention is determined by
\beq 
a_r = \left\{ \begin{array}{cc} 0, & \ \text{if} \ \exists \ i \ \text{s.t.} \ \norm{\sigma(V^T\tilde{x}+b_v) - 1/c_w\mu_{i}}_{\infty} < \theta, \\ 1, & \text{otherwise}. \end{array} \right.
\label{eq:decrule-addnewloc}
\eeq

When $a_r = 1$, it indicates that the attention is to be shifted. In this design, the location $i_s$ that the attention is shifted to is given by
\beq 
i_s = \text{argmax}_i \norm{\sigma(V^T\tilde{x}+b_v) - 1/c_w\mu_{i}}_{\infty}.
\label{eq:indsel-aftsat}
\eeq

The attention mechanism is initialized with the possible range of selections limited to just one location. The mechanism can expand this range to include other locations progressively if doing so could be beneficial. The decision to include new locations is specified by the same decision rule \eqref{eq:decrule-addnewloc}. This ensures that the controller starts with a limited set of memory locations and increases this set only when required. This can avoid unwanted jumps in attention, which could be the case with a mechanism that uses only hard attention. 

\subsection{\it NN Output:} The learning system (NN) {\it modifies its output} using the information $h_o$ retrieved from the memory. For this memory interface, the NN output is modified by adding the output of the Memory Read to the output of the hidden layer as given below.
\beq \text{NN Output:} \ u_{ad} = - \hat{W}^T\left(\sigma(\hat{V}^T\tilde{x} + \hat{b}_v) + h_o\right) - \hat{b}_w. \label{eq:nnoutput} \eeq

We postulated and showed empirical evidence in our earlier work \cite{muthirayan2019memory} that such a modification improves the speed of learning by inducing the search in a particular direction, which facilitates quick convergence to a neural network that is a good approximation of the unknown function. 
For a detailed discussion on how memory augmentation improves learning we refer the reader to \cite{muthirayan2019memory}. The computed NN output is fed to the controller (Fig. \ref{fig:cwmem}) which then computes the final control input by Eq. \eqref{eq:claw}.

{\it NN Update Law}: 
\label{sec:genmann}
The NN update law, which constitutes the learning algorithm for the proposed architecture, is the regular update law for a two layer NN \cite{lewis1998neural}:
\[
\left[\begin{array}{c} \dot{\hat{W}} \\ \dot{\hat{b}}^T_w \end{array}\right] = C_w\left(\hat{\sigma} - \hat{\sigma}^{'}  \left(\hat{V}^T\tilde{x}+\hat{b}_v\right)\right)h_e - \kappa C_w\norm{e}\left[\begin{array}{c} \hat{W} \\ \hat{b}^T_w \end{array}\right]  ,\nonumber
\]
\beq 
\left[\begin{array}{c} \dot{\hat{V}} \\ \dot{\hat{b}}^T_v \end{array}\right] = C_v \left[\begin{array} {c} \tilde{x} \\ 1 \end{array} \right] h_e \left[ \begin{array}{c} \hat{W} \\ \hat{b}^T_w \end{array}\right]^T \hat{\sigma}^{'} - \kappa C_v\norm{e}\left[\begin{array}{c} \hat{V} \\ \hat{b}^T_v \end{array}\right].
\label{eq:NNupdate} \eeq
The variable $h_e$ in \eqref{eq:memorywrite} is problem specific and depends on the Lyapunov function (without the NN error term). 

%
%

\section{Simulation Results and Discussion}
\label{sec:disc}
In this section, we consider several scenarios and show that the controller that uses the proposed attention mechanism results in improved performance. We also provide a detailed discussion on how the attention mechanism proposed here improves the performance. 
Finally, we provide a brief comparative study of the two dynamic key design approaches. 

\subsection{Robot Arm Controller} 
\label{sec:robotarmsys}
We briefly discuss the control equations for a typical robot arm controller augmented by an external working memory. The dynamics of a multi-arm robot system is given by, as in \cite{lewis1996multilayer},
\beq
M(x) \ddot{x} + V_m(x,\dot{x})\dot{x} + G(x) + F(\dot{x}) = \tau,
\label{eq:robotarm}
\eeq 
where $x \in \mathrm{R}^n$ is the joint variable vector, $M(x)$ is the inertia matrix, $V_m(x,\dot{x})$ is the coriolis/centripetal matrix, $G(x)$ us the gravity vector, $F(\dot{x})$ is the friction vector and $\tau$ is the torque control input. Let $s(t)$ be the desired trajectory (reference signal), then the error in tracking the desired trajectory is
\beq e(t) = s(t) - x(t). \eeq

Define the filtered tracking error by
\beq r = \dot{e} + \Lambda e, \eeq
where $\Lambda = \Lambda^T > 0$. Then the system equations in terms of the filtered tracking error $r$, as given in \cite{lewis1996multilayer}, is
\beq M\dot{r} = -V_m r - \tau + f, \eeq
where $f = M(\ddot{s} + \Lambda \dot{e}) + V_m (\dot{s} + \Lambda e) + N(x,\dot{x})$ and $ N(x,\dot{x}) =  G(x) + F(\dot{x})$. Given $f$, it folows that the input to the NN, $\tilde{x} = \left[e,\dot{e},s,\dot{s},\ddot{s}\right]$. For this system, the control law, \beq \tau = -u = -u_{bl} - u_{ad} - v, \label{eq:clawrobotarm} \eeq
where $u_{bl} = -K_v r$,
\[v = -k_v(\norm{\hat{W}}_F + \norm{\hat{V}}_F + \norm{\mu}_F+ Z_{m})r, \] 
and $u_{ad}$ is the NN output as defined in \eqref{eq:nnoutput}. The NN update laws are the same as \eqref{eq:NNupdate}. The vector $h_e = r^T$ in this case.

\subsection{Two Link Robot Arm System}
The system matrices for a typical two-link planar robot arm system are given below.
\beq M(x) = \left[\begin{array}{cc} \phi + \rho + 2\psi\cos (x_2) & \rho + \psi\cos (x_2) \\  \rho + \psi\cos (x_2) & \rho \end{array}\right], \label{eq:mass} \eeq
\beq V_m(x,\dot{x}) =  \left[\begin{array}{cc} -\psi\dot{x}_2\sin (x_2) &  -\psi(\dot{x}_1 + \dot{x}_2)\sin (x_2) \\ \psi\dot{x}_1\sin (x_2) & 0 \end{array}\right], \label{eq:cor} \eeq
\beq N(x,\dot{x}) =  \left[\begin{array}{c} \phi\gamma\cos (x_1) + \psi \gamma \cos (x_1 + x_2) \\ \psi\gamma\cos (x_1 + x_2) \end{array}\right]. \label{eq:} \eeq 

In the examples we consider here, the initial masses of the two links are set as: $m_1 = 0.8 \ \text{Kg}, m_2 = 2.3 \ \text{Kg}$. Their arm lengths are set as: $l_1 = l_2 = 1 \ \text{m}$. The parameters in the matrices, in terms of the link masses and the link lengths are: $\phi = (m_1 + m_2)l^2_1 = 3.1 \ \text{Kg} \text{m}^2, \rho = m_2 l_2^2 = 2.3 \ \text{Kg} \text{m}^2, \psi = m_2 l_1 l_2 = 2.3 \ \text{Kg} \text{m}^2, \gamma = g/l_1 = 9.8 \ \text{s}^{-2}$. 

The control algorithm parameters are set as: $c_w = 3/4, \theta = 0.2, K_v = 20, k_v = 10, \kappa = 0, C_w = C_v = 10$. The number of hidden layers, $N = 10$. To start with the number of memory vectors is initialized to one, i.e., $n_s = 1$. The memory is allowed to include up to a maximum of $5$ memory locations. For the memory interface, dynamic representation based key is used. Later, we draw comparison between this approach and the dynamic state based key approach. In all the scenarios described below, the attention reallocation is on only during the initializing phase. 

\subsection{Illustration of Attention Reallocation}

In the scenario we consider here, masses undergo the following sequence of abrupt changes:
\begin{align}
& m_i \rightarrow 2m_i \ \text{at} \ t = 10, \  m_i \rightarrow \sqrt{2}m_i \ t = 20, \ \text{and} \nonumber\\
& m_i \rightarrow 1/\sqrt{2}m_i \ \text{at} \ t = 40
\end{align}

The link lengths are: $l_1 = 1, l_2 = 2$. The values of other parameters are as given above. In the first set of results the attention reallocation is on only during the initial phase (case 1). In the second set of results the reallocation mechanism is on throughout (case 2).

\begin{figure}
\center
\begin{tabular}{ll}
\includegraphics[scale = 0.225]{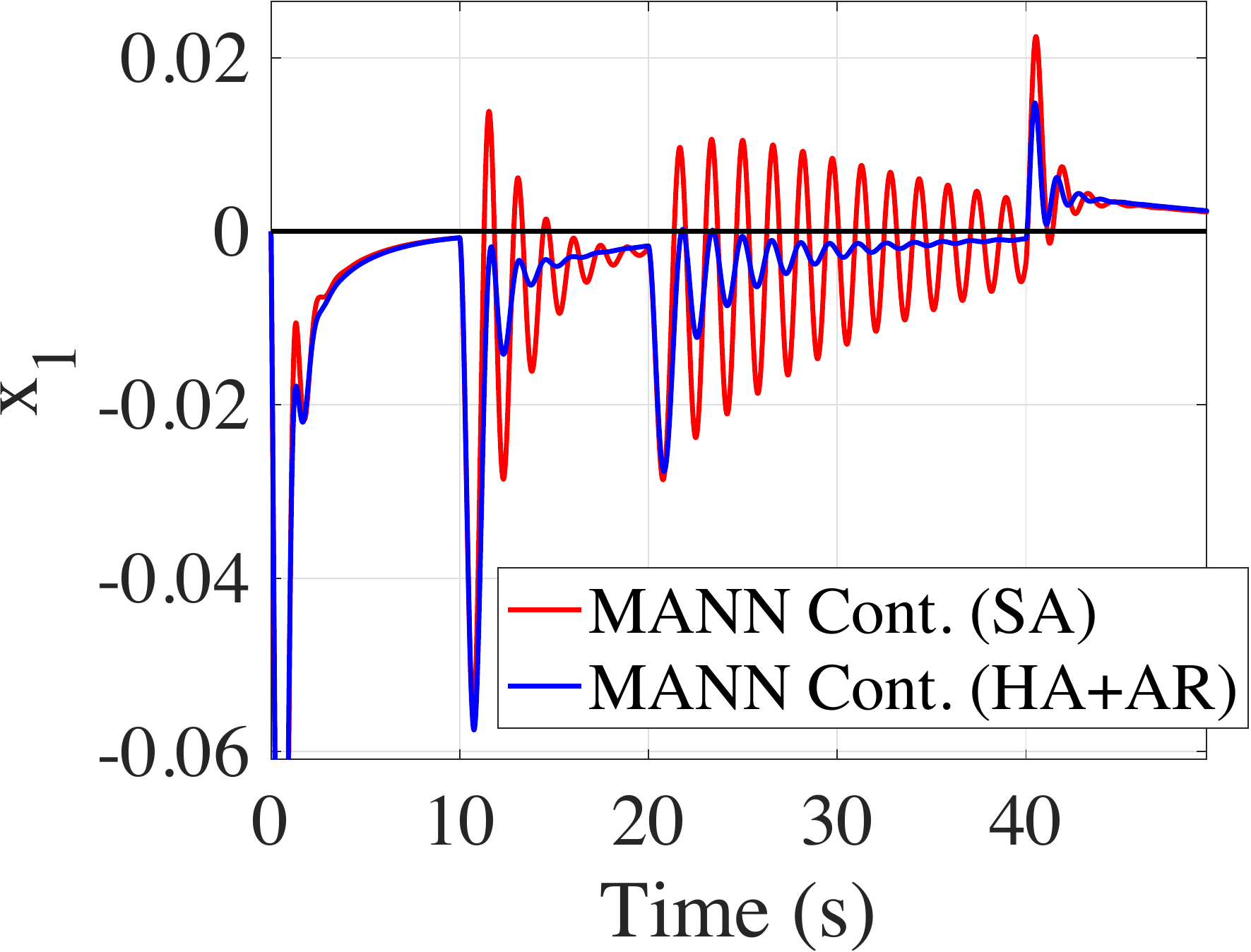} & \includegraphics[scale = 0.225]{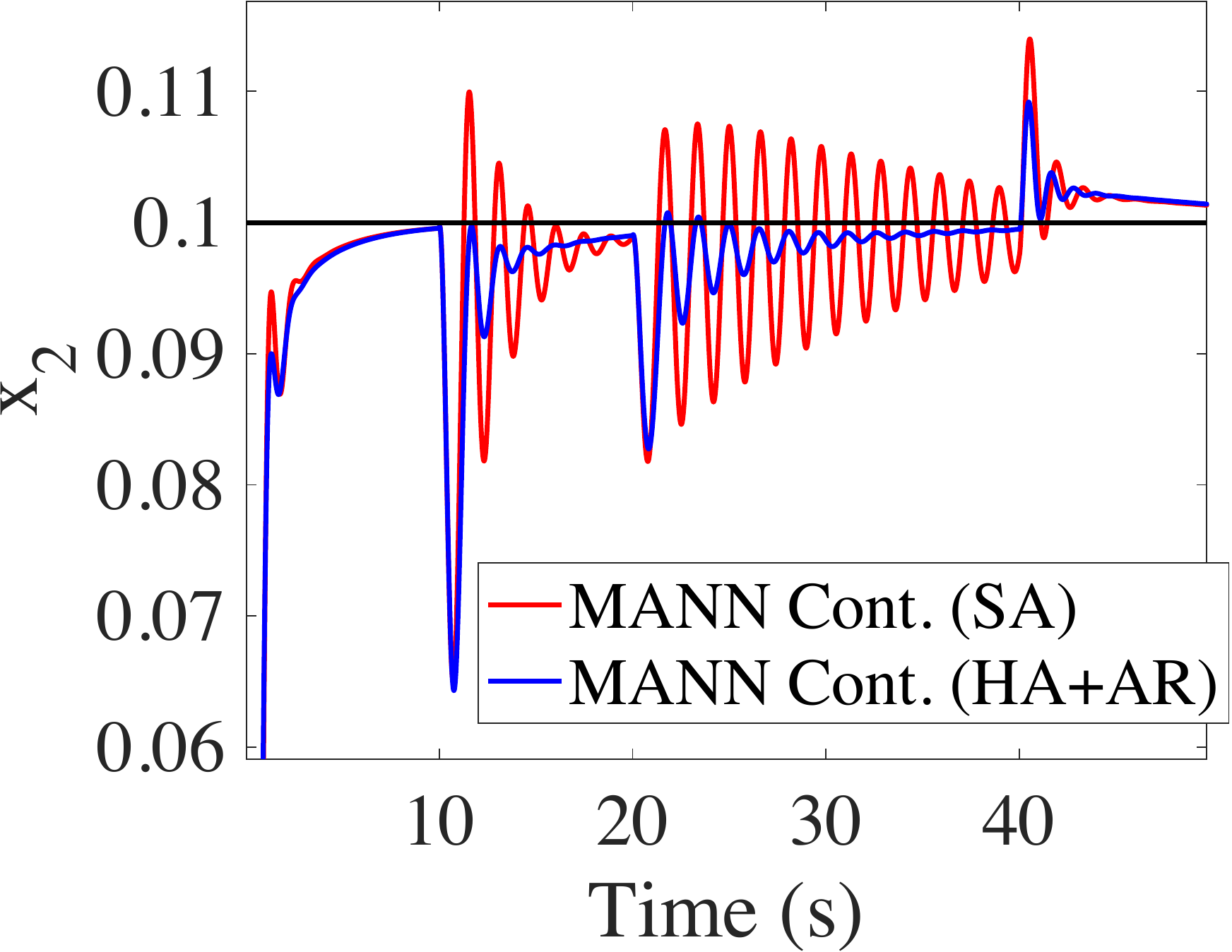} \\
\includegraphics[scale = 0.225]{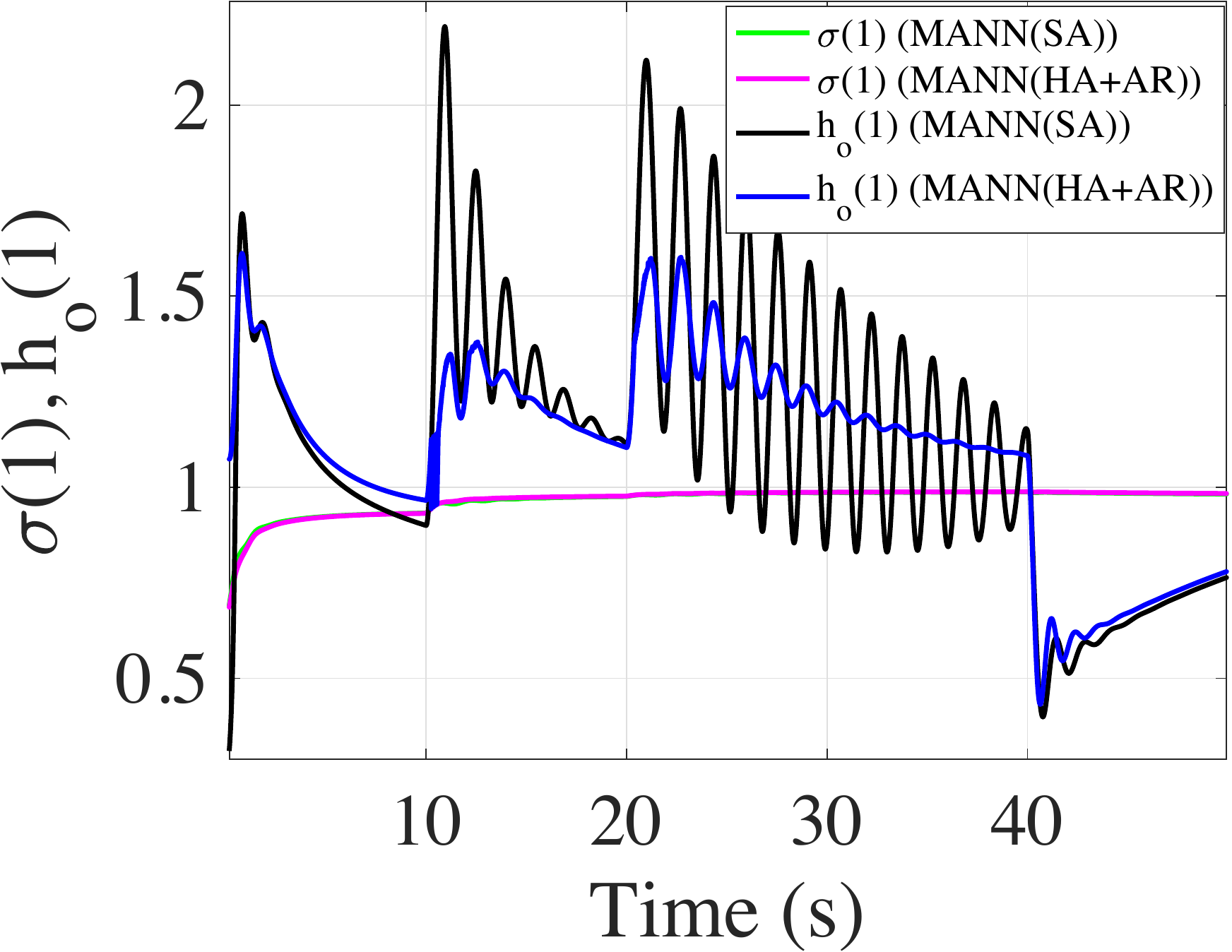} & \includegraphics[scale = 0.225]{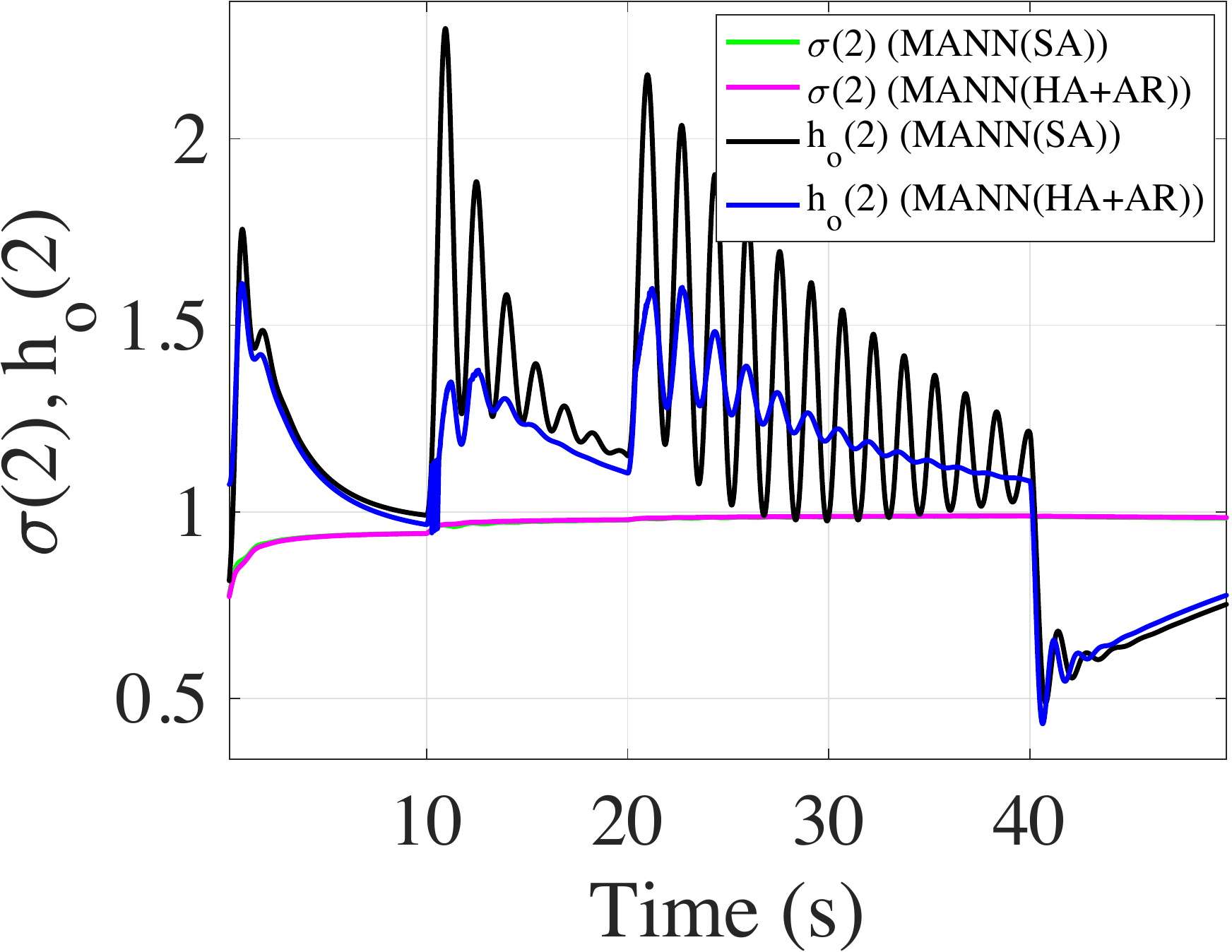}
\end{tabular}
\caption{Above: respone of first and second joint angle, $x_1$ and $x_2$, reallocation is on only during the initial period. Below: plot of $\sigma, h_o$.}
\label{fig:twolinkarm-0}
\end{figure}

\begin{figure}
\center
\begin{tabular}{ll}
\includegraphics[scale = 0.225]{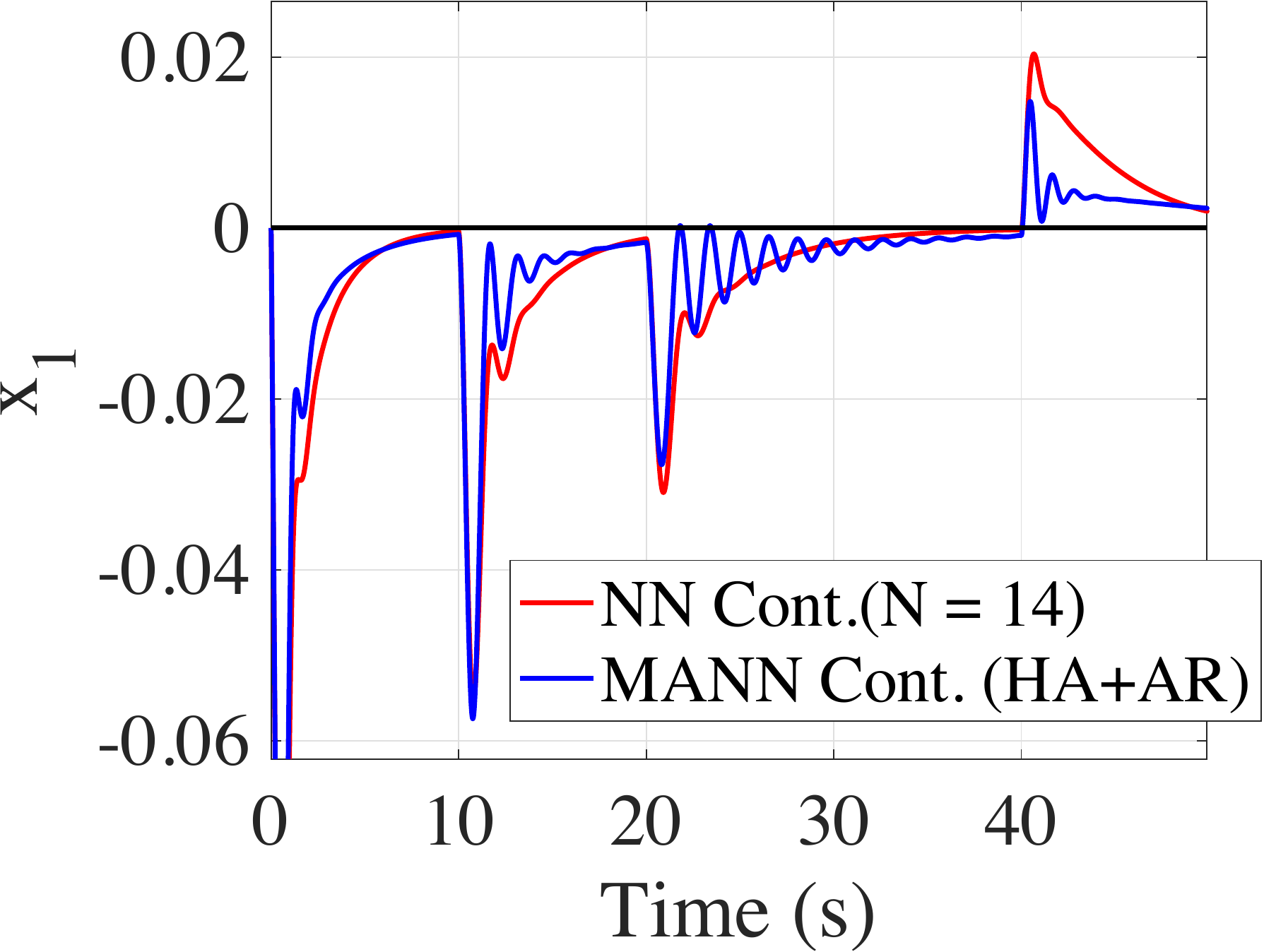} & \includegraphics[scale = 0.225]{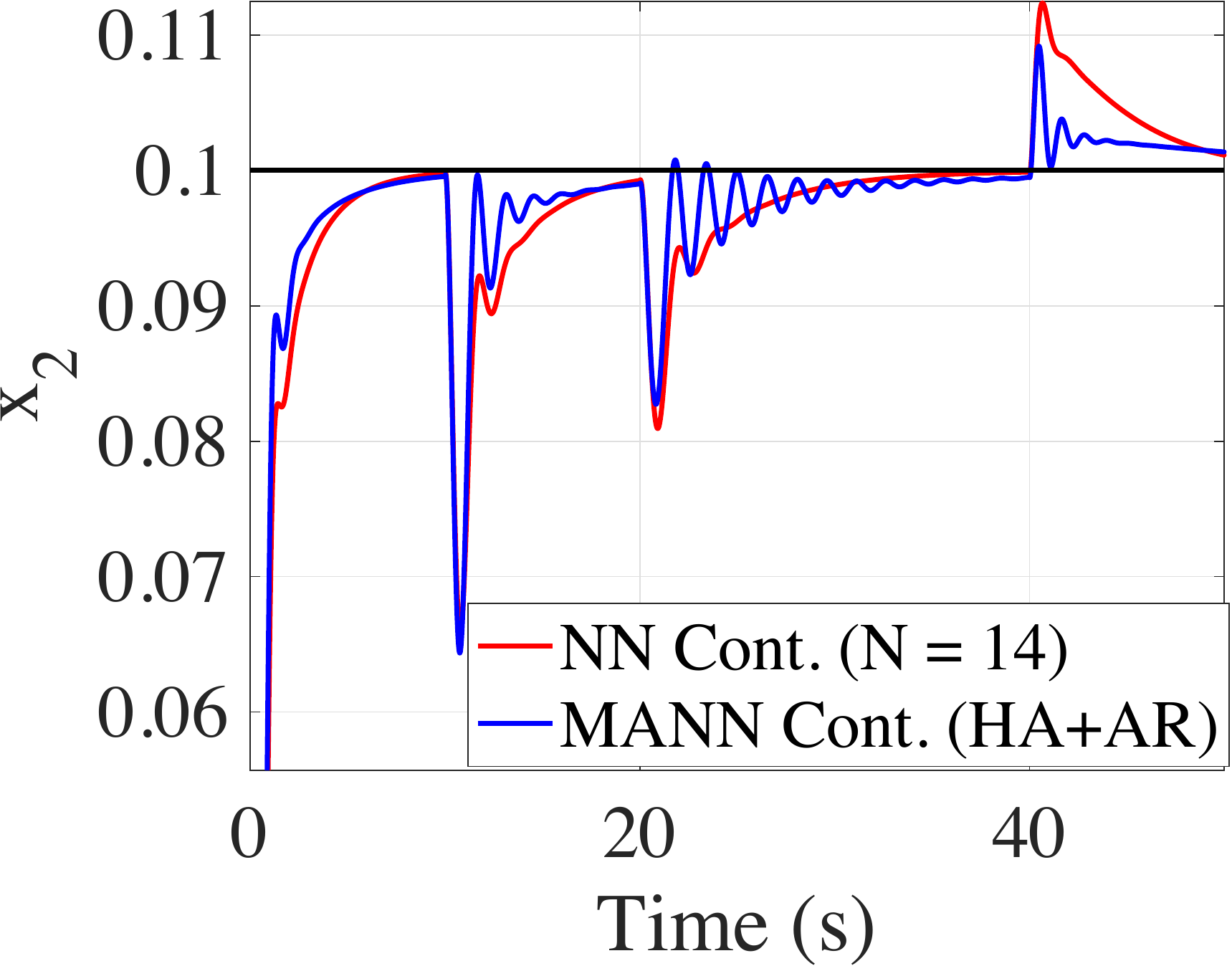} \\
\includegraphics[scale = 0.225]{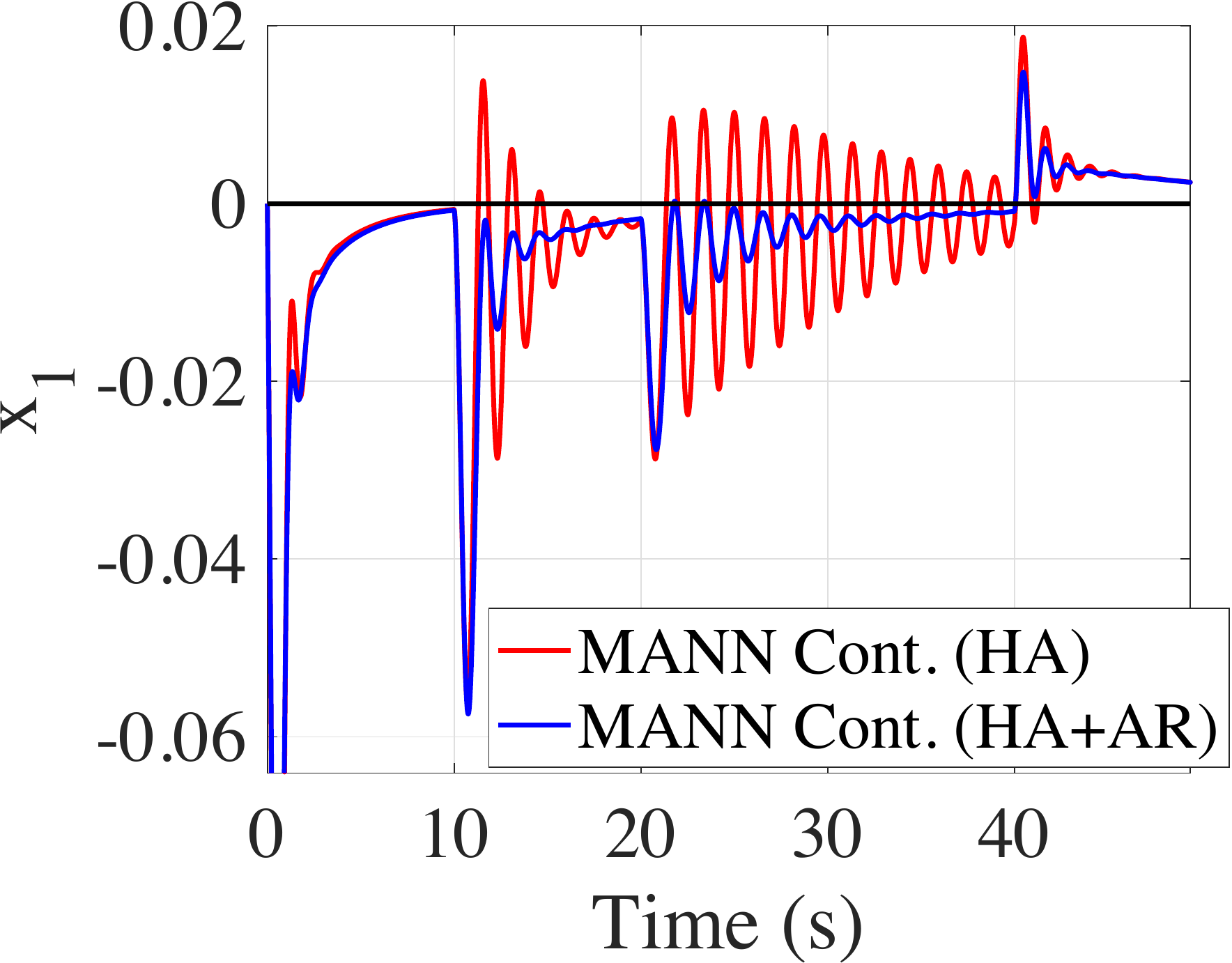} & \includegraphics[scale = 0.225]{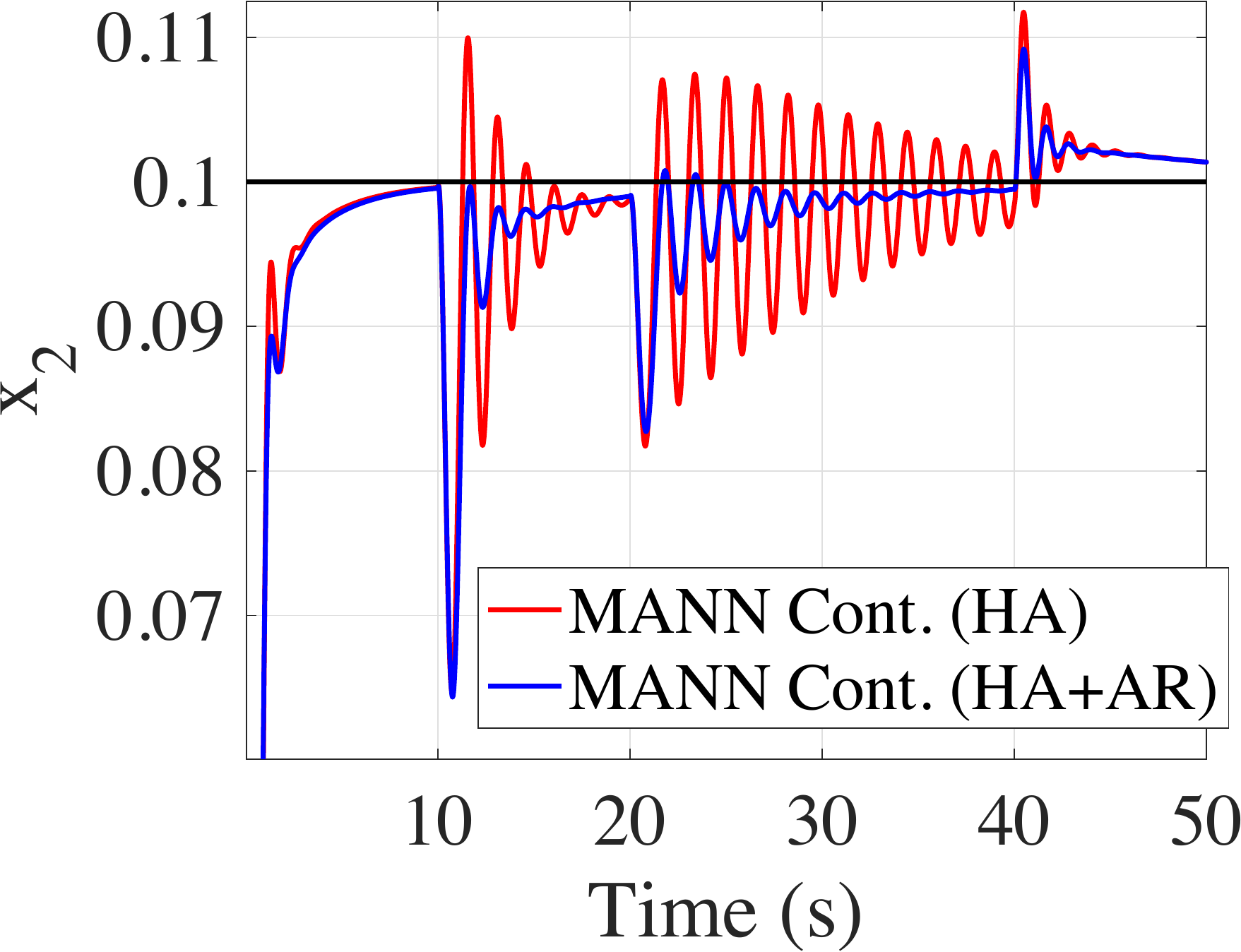} \\
\end{tabular}
\caption{Above: respone of first and second joint angle, $x_1$ and $x_2$. Comparison with a NN controller with $N = 14$. Below:  respone of first and second joint angle, $x_1$ and $x_2$. Comparison with the MANN controller that uses hard attention.}
\label{fig:twolinkarm-0-hard}
\end{figure}

{\it Case 1}: Fig. \ref{fig:twolinkarm-0} shows the response of joint angles for the controller that uses soft attention and the attention mechanism proposed here. Figure \ref{fig:twolinkarm-0-hard} shows the response of joint angles for the controller that uses hard attention and a regular NN controller with $N=14$ hidden layer neurons. We find that the responses for the controller that uses soft attention and hard attention show large and sustained oscillations after every abrupt change compared to the mechanism proposed in this work. From the figures it is clear that this is caused by the large change and the oscillations observed in $h_o$ after every abrupt change for these two controllers. 

The attention mechanism we proposed in this work shows an improved performance for the following reason. In the initial phase, the mechanism reallocates attention if the memory content deviates beyond the thershold $\theta$ and continues to reallocate if the deviations continue to occur. This continues till the number of memory locations grows up to $5$, the maximum that the memory can include. After this initial phase, the reallocation mechanism is turned off. But the information in the memory vector just before the reallocation is still retained and so the attention can shift back to this location if any subsequent jump exceeds the limit implied by this information. As a result, the Memory Read output is restricted from any deviation that exceeds the limit implied by the information in the memory contents even after the initial phase. This results in the diminished oscillations and thus the improved performance that we observe for the controller that uses hard attention and attention reallocation in the initial phase.

\begin{figure}
\center
\begin{tabular}{ll}
\includegraphics[scale = 0.225]{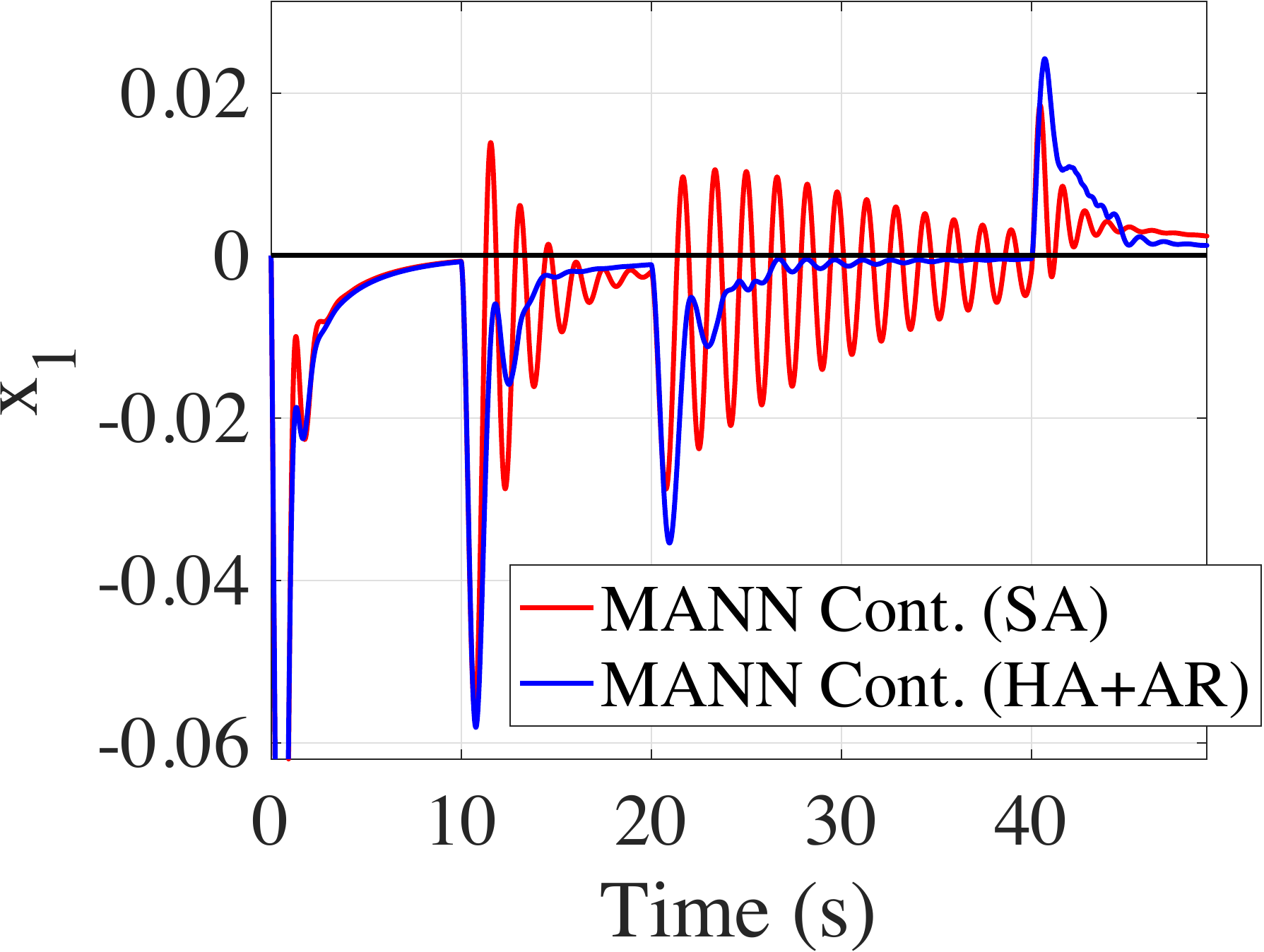} & \includegraphics[scale = 0.225]{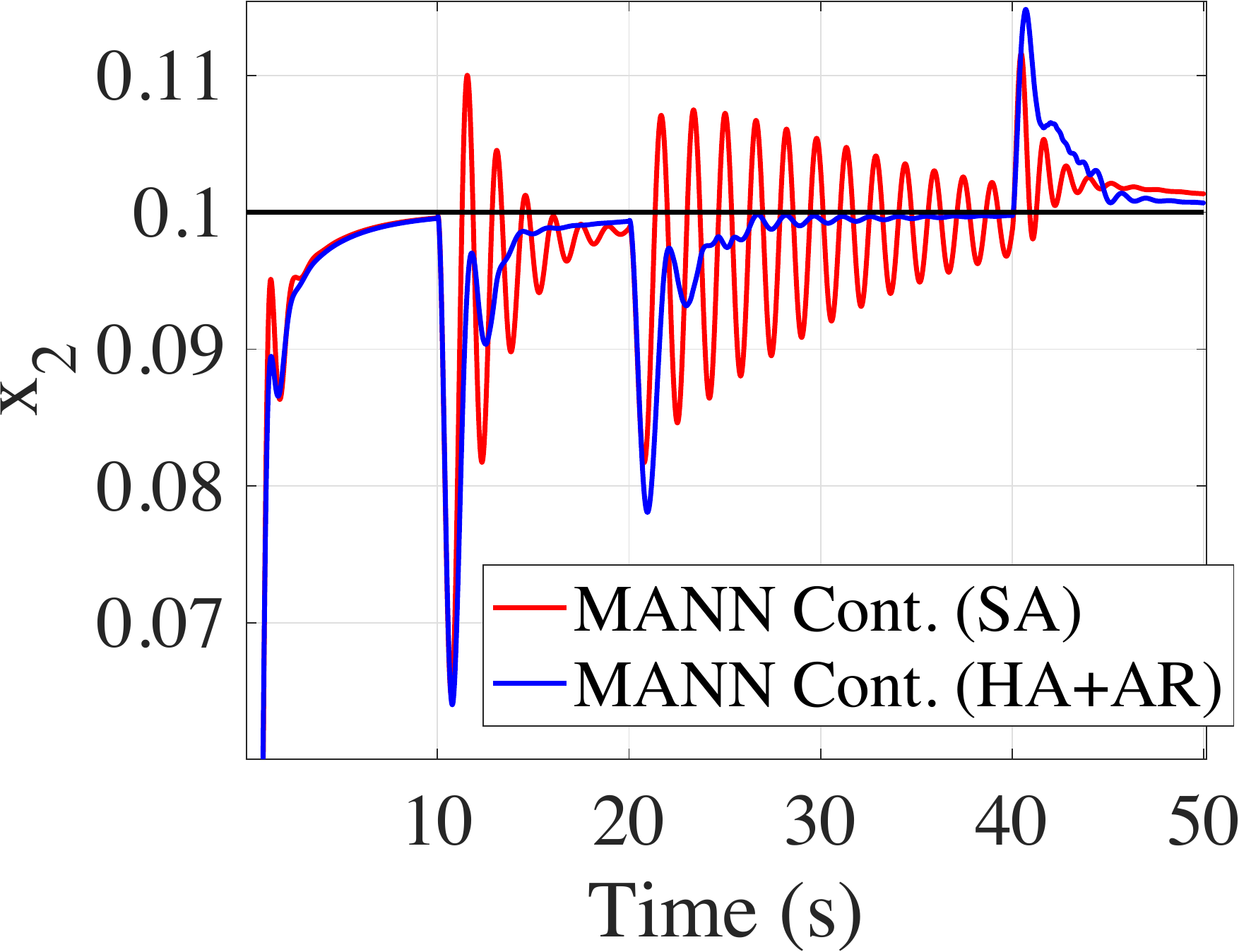} \\
\includegraphics[scale = 0.225]{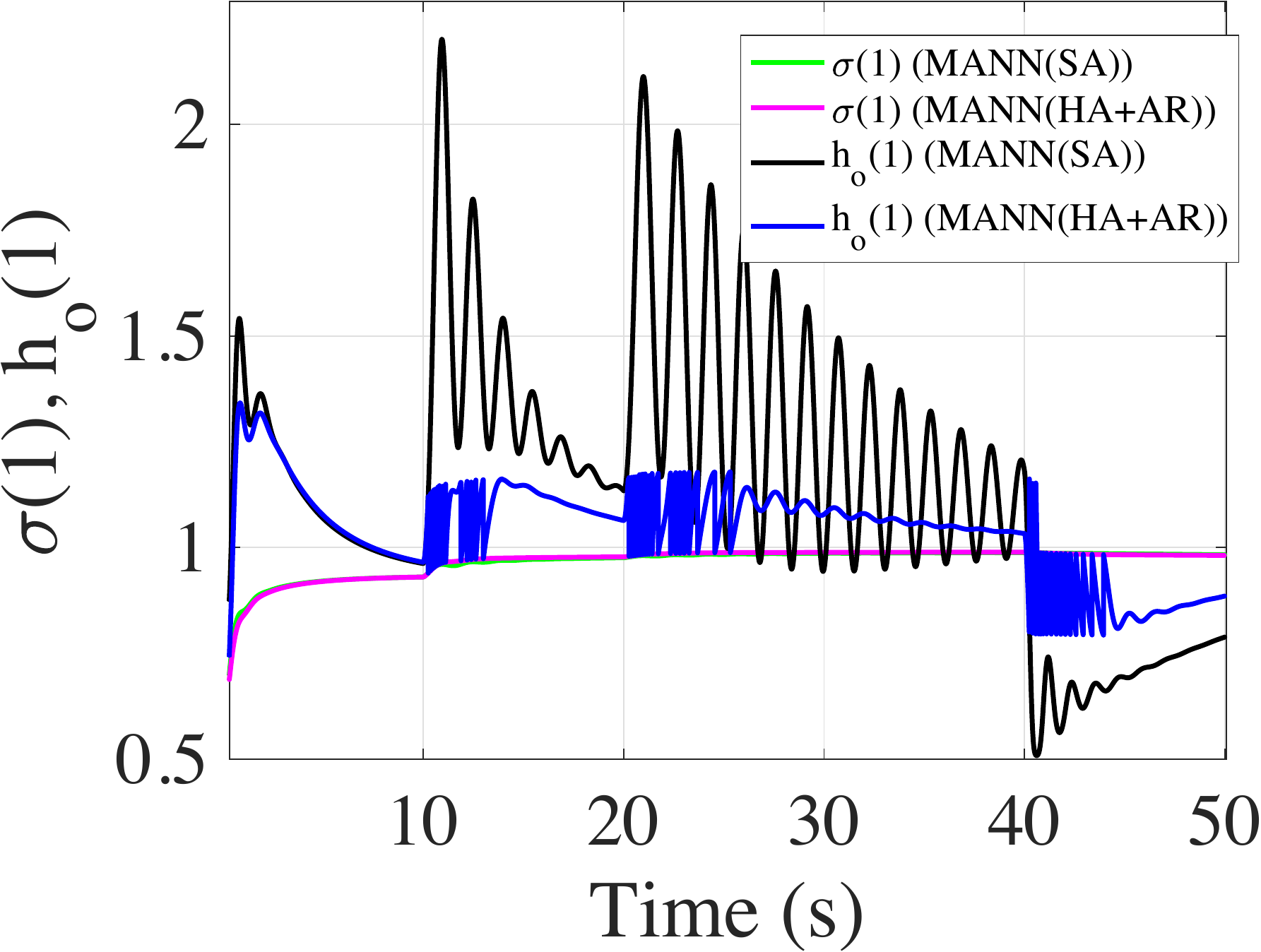} & \includegraphics[scale = 0.225]{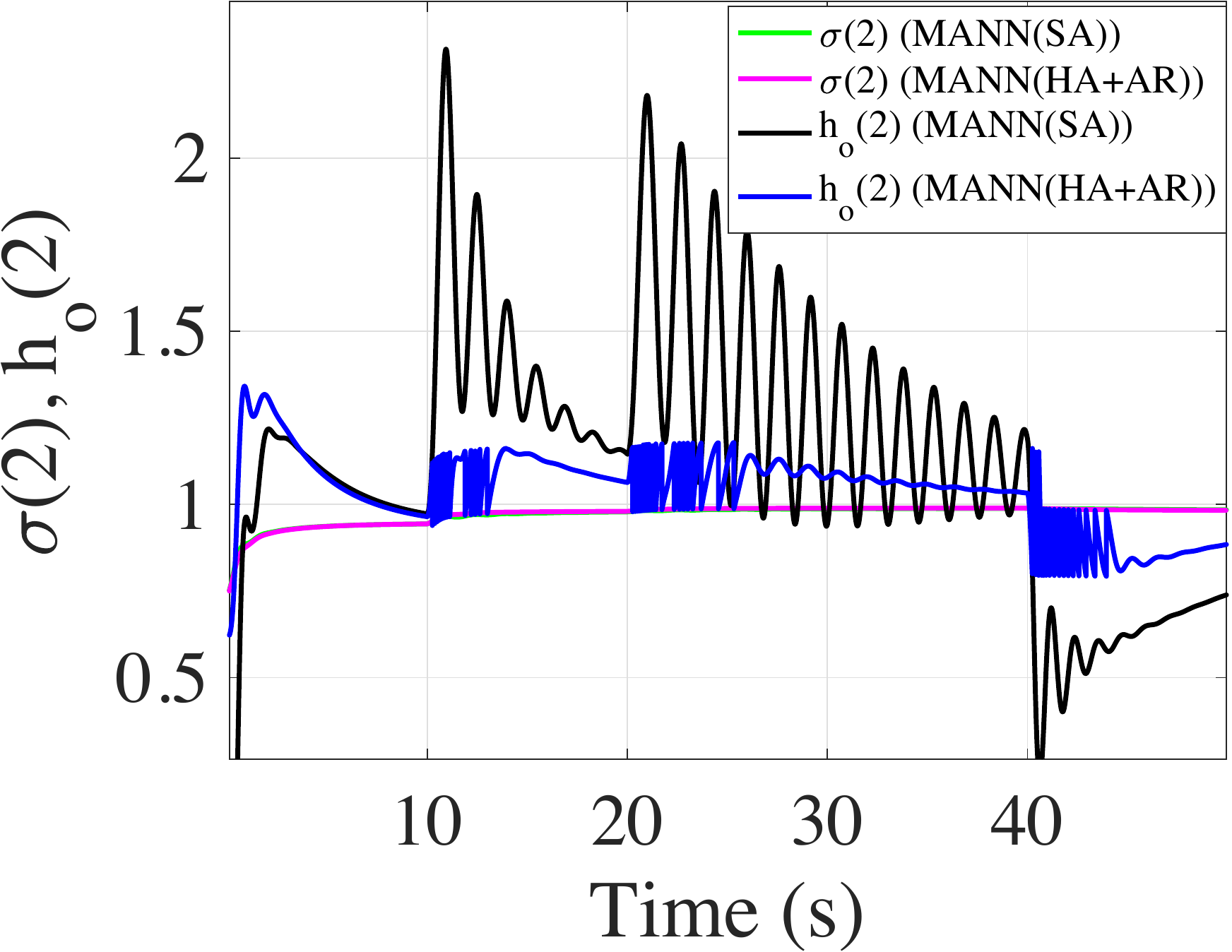}
\end{tabular}
\caption{Above: respone of first and second joint angle, $x_1$ and $x_2$, reallocation is on throughout. Below: plot of $\sigma, h_o$.}
\label{fig:twolinkarm-0-1}
\end{figure}

{\it Case 2}: Fig. \ref{fig:twolinkarm-0-1} shows the response of joint angles for the controller that uses attention reallocation throughout. The response shows almost negligible oscillations unlike the previous case. This is clearly attributable to the reallocation mechanism being active throughout, which restricts the deviation of the memory contents much more than the previous case, as is evident from the  $h_o$ and $\sigma$ plots shown in Fig. \ref{fig:twolinkarm-0-1}.

But, we observe that the initial peak after every abrupt change exceeds that of the controller that uses soft attention and the peak for the same controller in the previous case. This is observed because the quick correction provided by the second update term (or the third term) in the Memory Write equation \eqref{eq:memorywrite}, which necessarily results in a deviation, is restricted more than necessary in this case. Thus, having the attention reallocation on throughout reduces the oscillations but exaggerates the initial peak just after the abrupt change.

In the following, we present many scenarios to illustrate the effectiveness of the mechanism we proposed in this work. In all these scenarios, the major contributing factor to why the mechanism we proposed performs better is the explanation that we provided in this section.

\subsubsection{Scenario $1$} 

In this scenario, the values of the parameters are as given above. The attention reallocation is active only during the initial phase. It is set this way in all the scenarios we consider below. 

The command signals that the two arm angles have to track are: $s_1 = \text{sin}(0.5 t), s_2 = 0$ and the masses undergo the following sequence of abrupt changes:
{\small 
\begin{align} 
m_i & \rightarrow \sqrt{2} m_i \ \text{at} \ t = 5, \ m_i \rightarrow \sqrt{2} m_i \ \text{at} \ t = 25, \nonumber \\
m_i & \rightarrow \sqrt{2.5} m_i \ \text{at} \ t = 50 , \ m_i \rightarrow 0.63 m_i \ \text{at} \ t = 75, \nonumber \\
m_i & \rightarrow \sqrt{0.5} m_i \ \text{at} \ t = 90 , \ m_i \rightarrow \sqrt{0.5}  m_i \ \text{at} \ t = 110, \nonumber\\
m_i & \rightarrow \sqrt{0.1} m_i \ \text{at} \ t = 130 , \ m_i \rightarrow \sqrt{10}  m_i \ \text{at} \ t = 150, \nonumber\\
m_i & \rightarrow \sqrt{2} m_i \ \text{at} \ t = 170 , \ m_i \rightarrow \sqrt{5}  m_i \ \text{at} \ t = 190, \nonumber \\
m_i & \rightarrow \sqrt{0.2} m_i \ \text{at} \ t = 210 , \ m_i \rightarrow \sqrt{0.5}  m_i \ \text{at} \ t = 230, \nonumber \\
m_i & \rightarrow \sqrt{0.1} m_i \ \text{at} \ t = 250 , \ m_i \rightarrow \sqrt{10}  m_i \ \text{at} \ t = 270, \nonumber \\
m_i & \rightarrow \sqrt{2} m_i \ \text{at} \ t = 290 , \ m_i \rightarrow \sqrt{5}  m_i \ \text{at} \ t = 310.
\label{eq:jumps-sc1}
\end{align}}

In the results below, we compare the performace of this controller with the MANN controller discussed in our earlier work \cite{muthirayan2019memory} (uses soft attention) and an equivalent NN controller. An equivalent NN controller is a controller that has the same number of parameters as the MANN controller, where the parameter count for the MANN controller includes the size of the memory, which is $n_s\times N$. For the robot arm system considered in this section, the NN controller that is equivalent to a MANN controller with $N = 10$ hidden layer neurons and $n_s = 5$ number of memory vectors is a network with $N = 14$ number of hidden layer neurons.

Table \ref{tab:twolinkarm-1} provides the values of the sample root mean square error (SRMSE) for the two joint angles. The SRMSE values clearly indicate that the MANN controller version that uses hard attention (HA) and attention reallocation (AR) outperforms the other controllers by a notable margin.



\begin{table}[h]
\centering
\caption{Robot Arm System, SRMSE $\times 10^3$, Scenario $1$}
\begin{tabular}{|c|c|c|}
\hline
Joint Angle & 1 & 2  \\
\hline
NN Cont. (N = 14) & 10.2 & 4.4 \\
\hline
MANN Cont. (soft att., N = 10) - I & 8.8 & 3.8 \\
\hline
MANN Cont. (this paper, N = 10) - II & 8.2 & 3.6 \\
\hline
\% Reduction (From I to II) &  7.3 \% & 5.3 \%\\
\hline
\end{tabular} 
\label{tab:twolinkarm-1} 
\end{table}  

\subsubsection{Scenario $2$} 
In this scenario, the command signal for the joint angle $2$, $s_2 = 0.1$, instead of $0$. The masses undergo the same sequence of changes as given in \eqref{eq:jumps-sc1}. The MANN controller that uses soft attention is provided with $n_s = 5$ memory vectors and the equivalent NN controller is provided with a NN that has $N = 14$ number of hidden layer neurons. The rest of the setting for this scenario is similar to that of scenario $1$. For lack of space, we do not provide the plots of the responses for this scenario. Table \ref{tab:twolinkarm-3} provides the values of the sample root mean square error (SRMSE) for the two joint angles. We note that the proposed MANN controller outperforms the other controllers by a notable margin.

\begin{table}[h]
\centering
\caption{Robot Arm System, SRMSE $\times 10^3$, Scenario $2$}
\begin{tabular}{|c|c|c|}
\hline
Joint Angle & 1 & 2  \\
\hline
NN Cont. (N = 14) & 10 & 5.2 \\
\hline
MANN Cont. (soft att., N = 10) - I & 9.4 & 4.9\\
\hline
MANN Cont. (this paper, N = 10) - II & 8.3 & 4.5\\
\hline
\% Reduction (From I to II) &  11.7 \% & 8.2 \%\\
\hline
\end{tabular}
\label{tab:twolinkarm-3}
\end{table}


\subsubsection{Scenario $3$}
In the scenarios considered so far the sequence of changes were periodic in nature. The scenario we consider here is the same as scenario $1$ except that the abrupt changes the masses undergo result in a monotonic increase of the mass values. More specifically, we consider the following sequence of abrupt changes:

\beq m^2_i \rightarrow m^2_i + 0.2 m^2_i(0), \ \text{after every} \ 20 \text{s}. \label{eq:jumps-sc4}\eeq

Table \ref{tab:twolinkarm-5} provides the values of the sample root mean square error (SRMSE) for the two joint angles. We note that, here too, the proposed MANN controller outperforms the other controllers by a notable margin. 

\begin{table}[h]
\centering
\caption{Robot Arm System, SRMSE $\times 10^3$, Scenario $3$}
\begin{tabular}{|c|c|c|}
\hline
Joint Angle & 1 & 2\\
\hline
NN Cont. (N = 14) & 5.7 & 2.6\\
\hline
MANN Cont. (soft att., N = 10) - I & 5.6 &  2.5  \\
\hline
MANN Cont. (this paper, N = 10) - II & 5.2 & 2.2 \\
\hline
\% Reduction (From I to II) & 7 \% & 12 \% \\
\hline
\end{tabular}
\label{tab:twolinkarm-5}
\end{table} 
%

\subsubsection{Scenario $4$}
In this scenario, the intial masses of the links are given by, $m_1 = 3$ and $m_2 = 2$. Rest of the setting considered here is the same as that of scenario $1$. The parameters for the interface and in the control and update laws are also equal to the values used for scenario $1$. 
Table \ref{tab:twolinkarm-6} provides the values of the sample root mean square error (SRMSE) for the two joint angles. The SRMSE values indicate that the proposed controller is only marginally better than the controller that uses soft attention in this scenario. We reported this scenario to illustrate the point that the improvements are scenario dependent. 


\begin{table}[h]
\centering
\caption{Robot Arm System, SRMSE $\times 10^3$, Scenario $4$}

\begin{tabular}{|c|c|c|}
\hline
Joint Angle & 1 & 2  \\
\hline
NN Cont. (N = 14) & 12.0 & 3.6 \\
\hline
MANN Cont. (soft att., N = 10) - I & 10.2 & 3.1 \\
\hline
MANN Cont. (this paper, N = 10) - II & 10.0 & 3.0 \\
\hline
\% Reduction (From I to II) & 2 \% & 3 \%\\
\hline
\end{tabular} 
\label{tab:twolinkarm-6}
\end{table}

\subsubsection{Scenario $5$}
In this scenario, the setting is similar to scenario $1$ except that the link lengths are different. The links lengths are: $l_1 = 1$ and $l_2 = 2$. The command signals are: $s_1 = 0$ and $s_2 = \sin(0.5t)$. 
Table \ref{tab:twolinkarm-7} gives the SRMSE values for the three controllers. Here too, we observe that the controller that uses the attention mechanism proposed achieves a performance that is better compared to soft attention.


\begin{table}[h]
\centering
\caption{Robot Arm System, SRMSE $\times 10^3$, Scenario $5$}

\begin{tabular}{|c|c|c|}
\hline
Joint Angle & 1 & 2  \\
\hline
NN Cont. (N = 14) & 11.0 & 7.6 \\
\hline
MANN Cont. (soft att., N = 10) - I & 9.8 & 7.4 \\
\hline
MANN Cont. (this paper, N = 10) - II & 9.3 & 6.8 \\
\hline
\% Reduction (From I to II) & 5 \% & 8 \%\\
\hline
\end{tabular} 
\label{tab:twolinkarm-7}
\end{table}

\subsection{Comparison with Hard Attention} 

Here, we report several scenarios where the controller that uses the proposed attention mechanism is noticeably better than the controller that uses hard attention. 

First, we consider scenario $6$. In this scenario the masses undergo the same set of abrupt changes as given in scenario $1$. The arm lengths of the robot arm are: $l_1 = 1$, $l_2 = 2$. The command signals are: $s_1 = 0, s_2 = 0.1$. The threshold for attention reallocation, $\theta = 0.25$. 
Table \ref{tab:twolinkarm-1-hard} gives the SRMSE values for the two joint angles. We only report the SRMSE values of the signals after the first $10$s in order to report the values without the initial peak. We find that the mechanism proposed here shows a noticeable improvement in the performance. Similar responses are observed in scenario $4$ when the command signals are: $s_1 = 0$ and $s_2 = 0.1$.


Next, we consider scenario $5$. 
Table \ref{tab:twolinkarm-1-hard-1} gives the SRMSE values for the two joint angles. The error values are reported for the responses starting from $10$s to report the values without the initial peak. Clearly, the SRMSE values for the controller that uses the proposed attention mechanism show a noticeable improvement relative to the controller that uses hard attention.  

We find similar performance improvements in scenarios $1$ and scenario $4$ when the link lengths are set as: $l_1 = 1$, $l_2 = 2$. In scenarios $1, 2$ and $3$, the responses for the two controllers turned out to be similar. In scenario $4$, we found the controller that uses hard attention to be marginally superior. 

\begin{table}[h]
\centering
\caption{Robot Arm System, SRMSE $\times 10^3$ ($t > 10$), Scenario $6$}
\begin{tabular}{|c|c|c|}
\hline
Joint Angle & 1 & 2  \\
\hline
NN Cont. (N = 14) & 9.4 & 5.7 \\
\hline
MANN Cont. (hard att., N = 10) - I & 7.8 & 4.9 \\
\hline
MANN Cont. (this paper, N = 10) - II & 7.3 & 4.5 \\
\hline
\% Reduction (From I to II) &  6.4 \% & 8.2 \%\\
\hline
\end{tabular} 
\label{tab:twolinkarm-1-hard}
\end{table} 

\begin{table}[h]
\centering
\caption{Robot Arm System, SRMSE $\times 10^3$ ($t > 10$), Scenario $5$}
\begin{tabular}{|c|c|c|}
\hline
Joint Angle & 1 & 2  \\
\hline
NN Cont. (N = 14) & 9.7 & 6.9 \\
\hline
MANN Cont. (hard att., N = 10) - I & 7.9 & 6.3 \\
\hline
MANN Cont. (this paper, N = 10) - II & 7.5 & 5.9 \\
\hline
\% Reduction (From I to II) & 5 \% & 6.3 \%\\
\hline
\end{tabular} 
\label{tab:twolinkarm-1-hard-1}
\end{table}

\subsection{Discussion on Key Design Methods}

In this section, we provide a comparative study of representation based key design and state based key design. 
Table \ref{tab:attcomp} gives the SRMSE values for the two key design approaches in scenarios $1$ and $4$ respectively. 
It is clear that the dynamic state based key design is better than dynamic representation based key design in scenario $4$ but is worser in scenario $1$. 
This suggests that a clear conclusion cannot be drawn on which key design is superior. 

%

\begin{table}[h]
\centering 
\caption{Robot Arm System, SRMSE $\times 10^3$}

\begin{tabular}{|c|c|c|}
\hline
Joint Angle & 1 & 2  \\
\hline
scenario $1$ && \\
\hline
MANN Cont. (soft att.) & 8.8 & 3.8 \\
\hline
MANN Cont. (dyn rep key) & 8.2 & 3.6 \\
\hline
MANN Cont. (dyn state key) & 8.5 & 3.7 \\
\hline
scenario $4$&& \\
\hline
MANN Cont. (soft att.)& 10.2 & 3.1 \\
\hline
MANN Cont. (dyn rep key) & 10.0 & 3.0 \\
\hline
MANN Cont. (dyn state key) & 9.5 & 2.9 \\
\hline
\end{tabular}
\label{tab:attcomp}
\end{table}

%
%
%

\section{Conclusion}

In this work, we proposed a much improved attention mechanism for working memory augmented neural network adaptive controllers. The attention mechanism we proposed is a combination of a hard attention and an attention reallocation mechanism. The attention reallocation enables the memory to reallocate attention to a different location when the information in the location it is reading from becomes less relevant. This shifts the attention to a new location which is initialized with the most relevant information and also retains the older information in the location prior to the shift. Thus, the memory is able to overcome the limitations of prior soft and hard attention mechanisms. We showed through extensive simulations of various scenarios that the attention mechanism we proposed in this paper is more effective than the other standard attention mechanisms. 


\bibliographystyle{IEEEtran}
\bibliography{Refs}

\end{document}